\documentclass[prl,preprint,11pt]{revtex4-2}
\usepackage{amsmath,amsfonts}
\pdfoutput=1 
\usepackage{graphicx}
\usepackage[utf8]{inputenc}
\usepackage{amsmath,amssymb}
\usepackage{textcomp}
\setcitestyle{super}

\begin{document}

\title{Transient Optoplasmonic Detection of Single Proteins with Sub-Microsecond Resolution}
\author{Martin D. Baaske}
\author{N. Asgari}
\author{D. Punj}
\author{Michel Orrit}
\affiliation{Huygens-Kamerlingh Onnes Laboratory, Leiden University, Postbus 9504, 2300 RA Leiden, The Netherlands}
\date{\today}
\maketitle

\textbf{Optoplasmonic methods capable of single protein detection so far rely on analyte immobilization in order to facilitate detection\cite{Ament2012,Zijlstra2012,Dantham2013,Rosman2013,Beuwer2015,Zhang2020}. These detection schemes, even if they facilitate transient single-molecule detection\cite{Baaske2014, Kim2016} via consequent formation and cleavage of chemical bonds, typically exhibit time resolutions on the order of milliseconds.  The need for analyte immobilisation is a direct consequence of the minuscule dimensions of plasmonic near fields typically providing sub-attolitre-sized detection volumes which in turn demand sub-microsecond temporal resolution for the direct detection of proteins in motion. Here we show that such temporal resolution can indeed be achieved. We demonstrate the observation of single proteins as small as Hemoglobin (molecular weight: $\mathbf{64}$ kDa) as they traverse plasmonic near fields of gold nanorods and interact with their surface, all while maintaining signal-to-noise ratios larger than $5$ and an unprecedented temporal resolution well below microseconds. This method enables the label-free observation of single-molecule dynamics on previously unaccessible timescales.}

In order to achieve such resolution we have improved the confocal system we had previously used to detect single metal nanoparticles in Brownian motion\cite{Baaske2020}. Specifically we achieve gains in signal-to-noise ratio via optimization of incident and analyzed polarization, probe wavelength and Gouy phase for individual plasmonic gold nanorods. The setup is shown in Fig.\,1a and was specifically designed to maintain linear polarization states.  
For each NR we determine its orientation via rotation of linear incident and analyzer polarisation and obtain its white-light scattering spectra, which allow us to determine whether the overlap of the NRs spectrum with the tuning range of our probe laser $(785\pm20)$\,nm is suited. We then align the incident and analyzer polarisations parallel with the NR's axis and select the wavelength which exhibits the highest slope in the scattering spectra. The detected intensity $I_{det} \propto |E_r|^2 + |E_s|^2 + 2 |E_r||E_s|\cos(\Delta\phi) $ is a result of the interference between the scattered $E_s$ and reflected  $E_r$ fields projected onto the analyzer's axis and the phase difference between both fields is $\Delta \phi$. NRs with different scattering cross sections require different experimental parameters in order to maximize signal-to-noise. To address this requirement we utilize photothermal spectroscopy\cite{Gaiduk2010,Selmke2012,Adhikari2020,Chen2021} as a convenient probe of an individual NR's sensitivity. This is done by heating the nanorod with the intensity-modulated  $532$\,nm laser (modulation frequency $\nu_H=1.3$\,MHz) while probing it with the intensity-modulated wavelength-tunable probe laser (modulation frequency $\nu_P=1$\,MHz). Heating the NR gives rise to minute changes of the surrounding medium's refractive index which in turn causes the NR's localized surface plasmon resonance (LPSR) to shift. This shift is then detected as a change of $I_{Det}$ at the modulation frequency $\nu_H$.   
The modulation of the probe beam further enables us to directly determine the relative photothermal amplitude $A_{PT} = \frac{A(\nu_{H})}{A(\nu_{P})}$, where $A(\nu)$ denotes the root-mean-squared (rms) amplitude at the respective frequency $\nu$, with minimal contributions from $1/f$-noise sources. $A_{PT}$ is proportional to the relative change of intensity as a response to the photothermal refractive index change over the total detected intensity and therefore a direct measure of the NR's sensitivity, to be optimized by the adjustment of such experiment's parameters as probe wavelength, probe power and the NR's position in the focal volume. 
For our measurements we use nanorods with dimensions of $25\times80$ and $40\times110$\,nm and typical LSPR's of $1.6$\,eV ($775$\,nm).

\begin{figure}[htbp]
\centering
\includegraphics{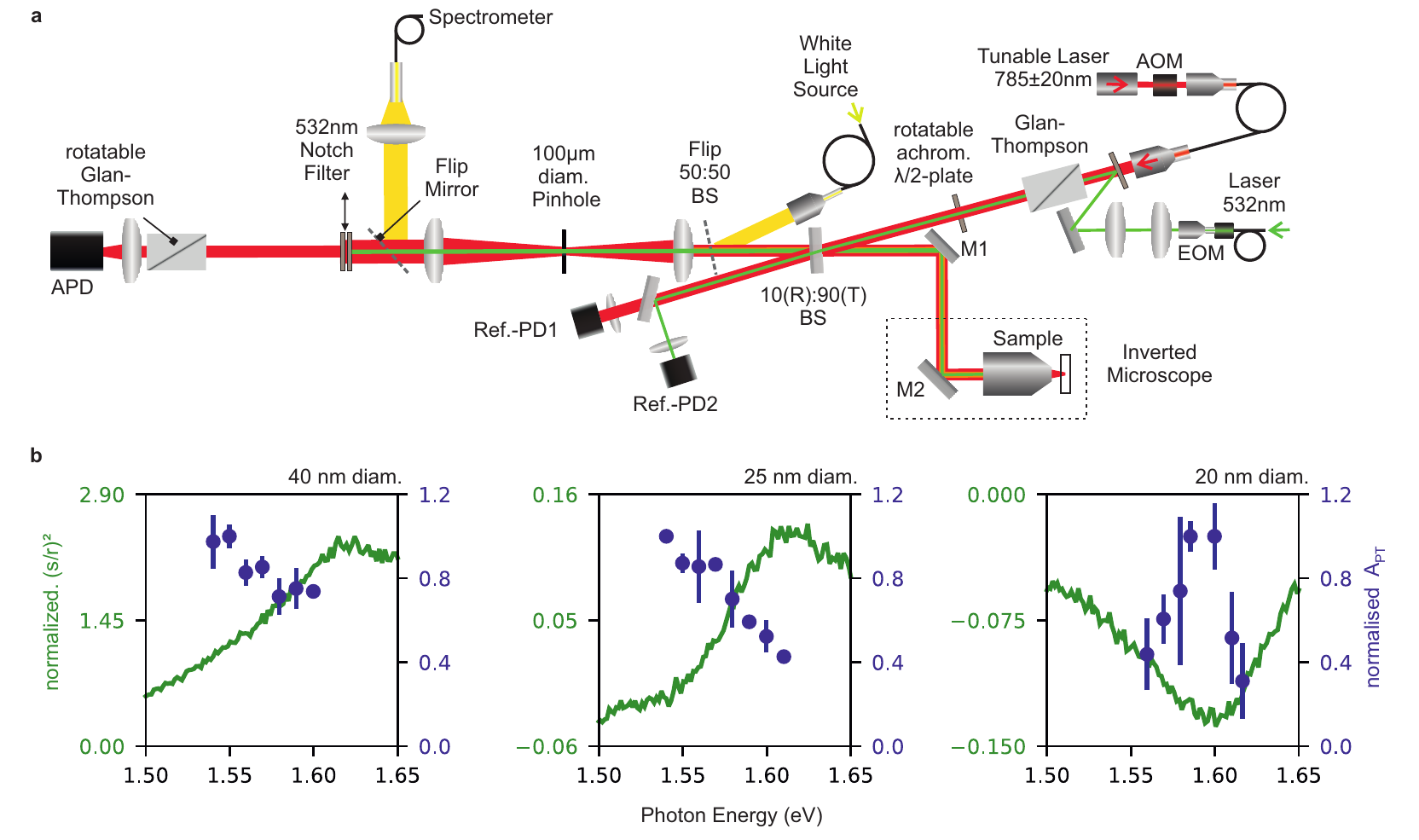}
\caption{Experimental setup with photothermal signal optimization: Panel \textbf{a} shows the optical setup. Panel \textbf{b} shows sections of white-light scattering spectra normalized to the reflection on the glass slide ($(s/r)^2$, green line) and the corresponding relative photothermal amplitude $A_{PT}$ values (blue dots) obtained while varying the tunable laser's wavelength for NRs with various diameters.}
\end{figure}    

For NRs with $\approx40$\,nm diameter the on-resonance scattering coefficient $s=E_s/E_0$, where $E_0$ is the incident field, significantly exceeds the reflection coefficient $r=E_r/E_0$ of the glass-water interface and the interference term is therefore negligible. In this case high $A_{PT}$ values are found at a wavelength coinciding with the flanks of the NR's LSPR spectrum i.e. where the slopes of the Lorentzian are highest (see Fig.1\,b, left).  For the $20-25$\,nm diameter NRs $s\approx r$ and high $A_{PT}$ values are found closer to the LSPR center (see Fig.1\,b, center and right). We find that the interplay between contributions of scattering cross section and phase changes upon heating is more complex and also dependent on the Gouy phase i.e. the NR's distance to the focal plane. In order to streamline the alignment process we follow a simple recipe for all NRs: We first center the NR in the focus (xyz) by maximizing the scattered intensity with crossed polarizers. Then we optimize $A_{PT}$ with parallel polarizers aligned along the NR's long axis by first tuning the wavelength and then adjusting the NR position along the focal axis (z).  

Single-molecule measurements are typically performed by recording intensity time traces of $10$\,ms length with a sampling rate of $100$\,MHz. These traces are then de-noised by applying a running $10$-point median filter. Specifically we detect Glucose Oxidase (GOx, molecular weight $\approx160$\,kDa) from \textit{Aspergillus niger} (Fig.\,2a) and Hemoglobin (Hem, $\approx64$\,kDa) from bovine blood (Fig.\,2b) molecules as they move through the NR's near field. These molecules produce transient shifts of the LSPR, which are recognized as changes of the detected intensity. These changes appear on the intensity traces as patterns of two different types: 1) Short spike-like perturbations (comp. Fig.\, 2a I-1, II-1 left, III-2 and 2b II-1, II-2, III-2) which we interpret as protein molecules moving directly in and out of the the near-field. 2) Level-transition patterns  (comp. Fig.\,2a I-1 and III-1 center and Fig.\,2b I, III-1 and IV-1) which we think are caused by protein molecules moving through the aqueous medium into the near field and then dwelling at the NR's surface due to attractive forces until the attractive potential is eventually overcome and they again move out of the near field. We also observed a few binding and unbinding events without their respective counterparts in the same trace. This suggests that sticking lasting longer than $10$\,ms occurs.
\begin{figure}[htbp]
\centering
\includegraphics{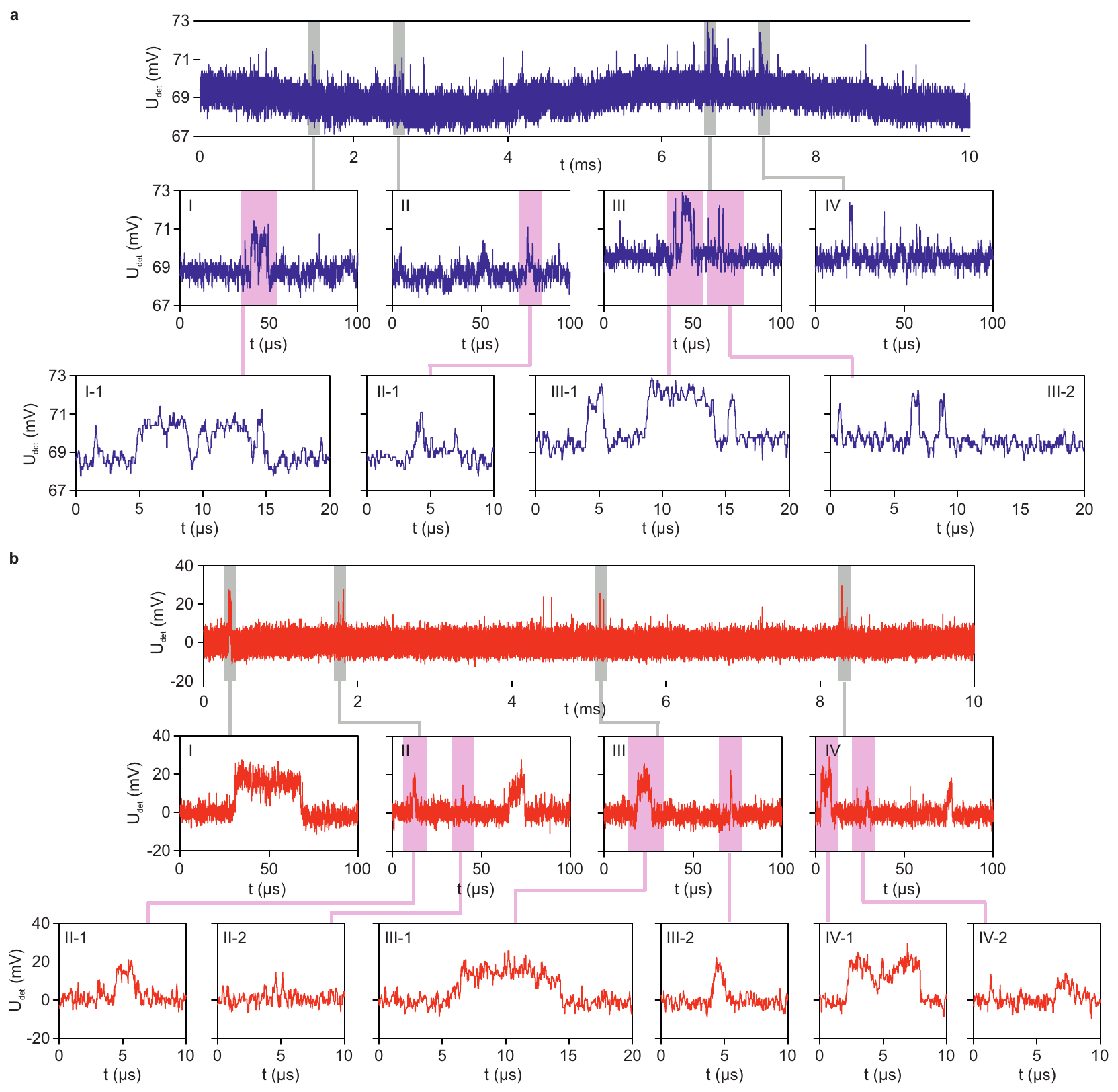}
\caption{Intensity perturbations due to GOx (blue) and Hem (red) molecules moving in a NR's near field. \textbf{a}) Trace (top) and shorter timescale subtraces (I - IV) showing perturbations caused by Glucose Oxidase molecules observed with a $25$\,nm diameter NR and a DC-coupled detector. \textbf{b}) Trace (top) and shorter timescale subtraces (I - IV) showing perturbations caused by Hemoglobin molecules observed with a $40$\,nm diameter NR and an AC-coupled detector. Most perturbations exhibit sub-\textmu s rise- and fall-times. Longer perturbations as \textbf{a} I-1 and III-1 center, \textbf{b} I, III-1 and IV-1) are likely caused by attractive interactions between analyte and NR. All traces were recorded with $\Delta t=10$\,ns and de-noised with a $10$-point median filter. Protein concentrations were $500$\,nM (\textbf{a}, GOx) and $30$\,nM (\textbf{b}, Hem)}
\end{figure}

In order to obtain statistics we count fluctuations of intensity traces as events if their amplitudes exceed $5\sigma$, where $\sigma$ is the standard deviation of the whole trace. This does not include rare single step events without counterpart in the same trace (GOx). In the case of (Hem) level transitions with durations longer than $0.1$\,ms were excluded from analysis in order to avoid artifacts due to the low-frequency cut-off ($1.5$\,kHz) of the AC-coupled detector.      
For each event we determine the maximum amplitudes $\Delta I_{max}$ and the following temporal properties: $\tau_{rise}$ ($\tau_{fall}$) the rise (fall) time i.e. the time it takes from the beginning (end) of the event to rise to half the maximum and the duration between these points i.e. the full duration at half maximum (FDHM). 
\begin{figure}[htbp]
\centering
\includegraphics{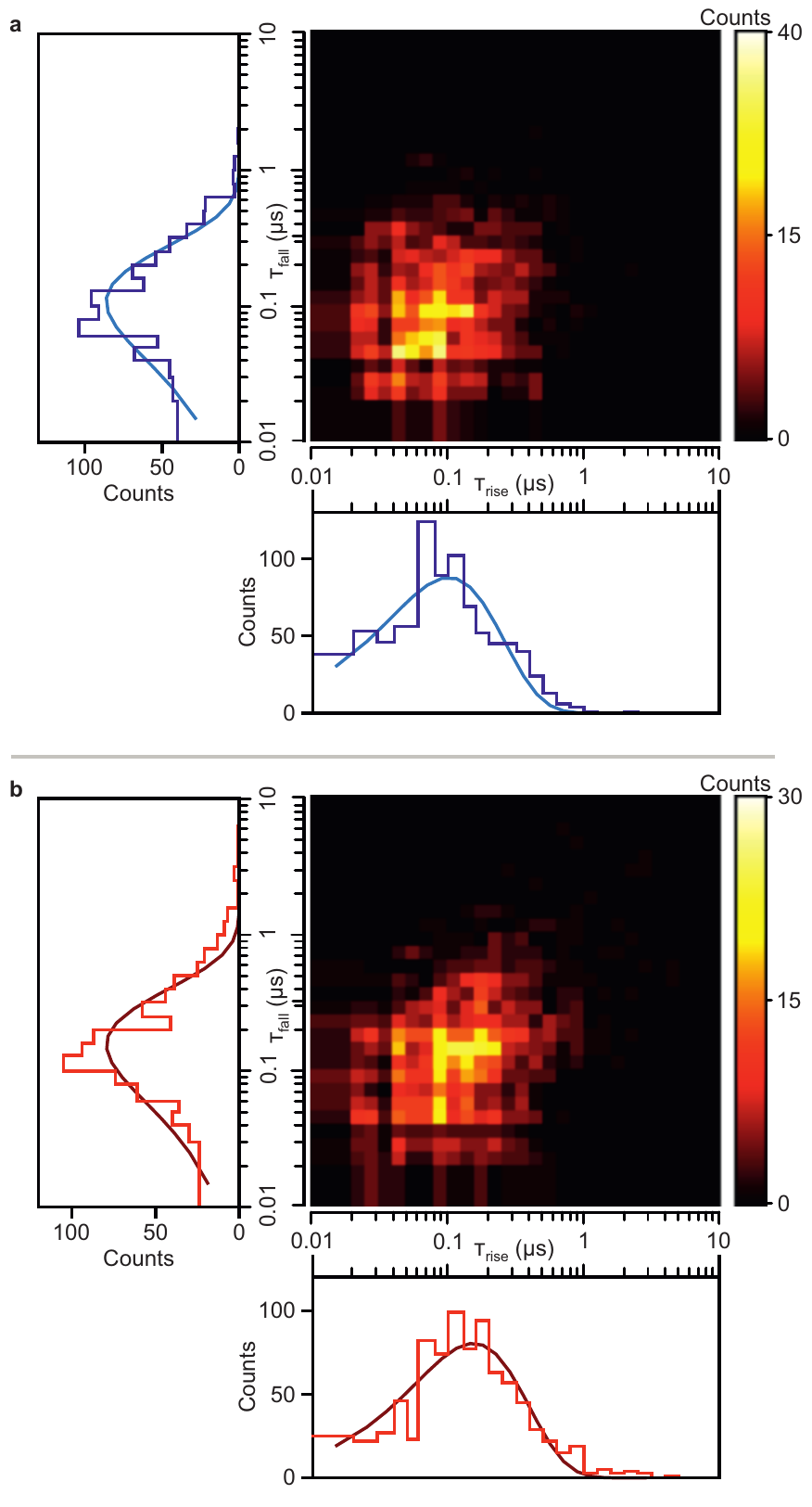}
\caption{Statistics of rise and fall times, characterizing the molecular diffusion in the abscence of sticking as well as immediately before and after sticking, for GOx (\textbf{a}) and Hem (\textbf{b}). Each panel (\textbf{a},\textbf{b}) shows  the $\tau_{rise}$ (bottom) and $\tau_{fall}$ (right) distributions alongside their respective 2D-histograms (center). Solid lines (\textbf{a}, light blue and \textbf{b}, dark red) in the distributions show fits to mono-exponential time distributions plotted on logarithmic scale of times.}
\end{figure}     

For both Hem and GOx we find rise- and fall-times in the range from $10$ to $1000$\,ns (see Fig.\,3a, b). All $\tau_{rise}$ and $\tau_{fall}$ distributions fit well to mono-exponential decays (i.e. to $f(t) = N \frac{t}{\tau} e^{-t/\tau}$ for the logarithmically spaced distributions) and we find rise (fall) decay times of $101\pm8$\,ns ($108\pm6$\,ns) for GOx and $153\pm11$\,ns ($155\pm11$\,ns) for Hem. Using these values as diffusion times for spheres with hydrodynamic radii of $4.45$\,nm (GOx)\cite{Courjean2009}, $3.11$\,nm (Hem)\cite{Arosio2002} we find rms displacements of $\approx6$\,nm (GOx) and $8$\,nm (Hem) which match well with the half decay length of optical near fields. The similarity between rise- and fall-time distributions suggests that entry and the exit processes of both proteins into and out of the NR's sensitive volume are subject to equivalent interactions. The 2D-histograms (Fig.\,3a and 3b, center) further show no apparent correlation between $\tau_{rise}$ and $\tau_{fall}$ for individual events further suggesting that entry and exit processes into and out of near field are mutually independent as we would expect for Brownian motion. The rise and fall times for GOx are shorter than the ones we find for Hem although GOx exhibits the higher molecular weight. We speculate that this difference is due to the (in comparison to the globular Hem) more anisotropic shape of GOx\cite{Wohlfahrt1999} and may well reflect an additional contribution from rotational diffusion. For GOx we estimate a rotational diffusion time in the order of $50$\,ns which falls well into our temporal resolution. Differences in the autocorrelation measurements for both molecules further support this hypothesis (see suppl. info. section S1). Nonetheless further work beyond the scope of this manuscript is needed to confirm this.
\begin{figure}[htbp]
\centering
\includegraphics{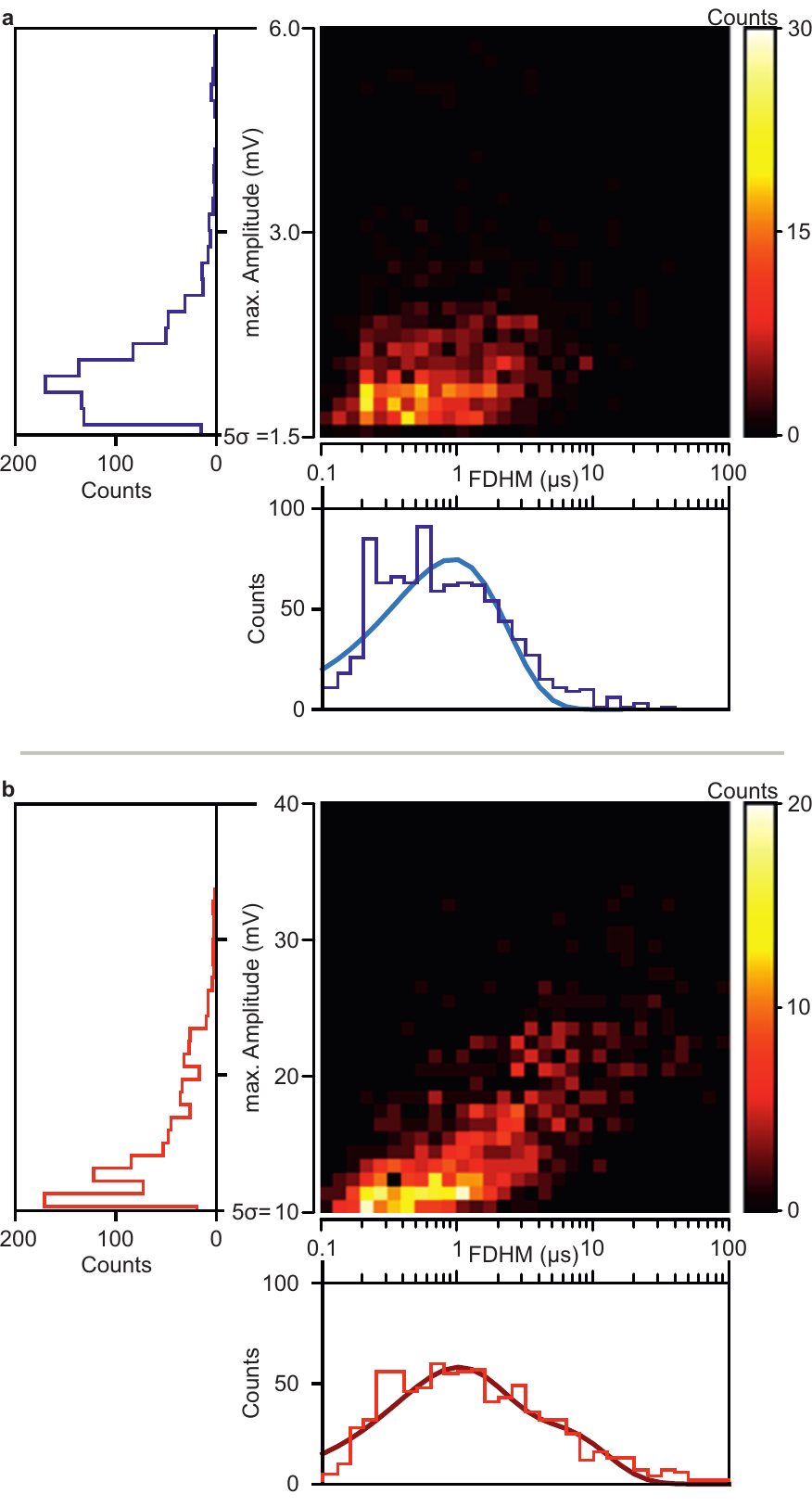}
\caption{Statistics of dwell times and amplitudes as obtained for GOx (\textbf{a}) and Hem (\textbf{b}). Each panel (\textbf{a},\textbf{b}) shows  the FDHM (bottom) and maximum amplitude (right) distributions alongside their respective 2D-histograms (center). Solid lines in the FDHM distributions show fits to mono- (\textbf{a}, light blue) and bi-exponential time distributions, plotted on logarithmic scale of times (\textbf{b}, dark red). }
\end{figure}
We also find similar FDHMs distributions for both proteins species. These times represent the dwell time of individual proteins in the NR's sensitive volume and are centered around $\approx1$\,\textmu s and distributed over ranges from $100$\,ns up to few tens of microseconds i.e. significantly longer than the rise- and fall-times alone (see Fig.\,3b). 
Here the stretch towards long timescales is likely due to attractive interactions between the proteins and the NR's surface, presumably unspecific sticking. The FDHM distribution of Hem (Fig.4\, b, bottom) is more stretched towards long times than the one of GOx (Fig.4\, a, bottom). Specifically we find that the FDHM distribution for GOx fits well to a mono-exponential decay as expected for a Langmuir adsorption process governed by a single time constant for which we find $\tau=0.9$\,\textmu s from the respective fit. For Hem this is not the case as the single rate fit clearly deviates from the data (not shown). This suggests contributions from multiple processes with different rates. Already for two rates ($\tau_1=0.83$\,\textmu s and $\tau_2 = 4.8$\,\textmu s) we find a much better match. This is rather unsurprising due to the larger set of fitting parameters. We think it is more likely that a broader spectrum of rates exists rather than just two. 
The maximum amplitude \textit{vs.} FDHM distribution for Hem (Fig.4\, b, center) also reveals a stronger correlation between larger amplitudes and longer times as compared with GOx (Fig.\,4a, center). 
We hypothesize the differences between Hem and GOx, specifically the existence of multiple rate constants could arise due to the following reasons: 1) Hemoglobin exhibits surface areas which possess different affinity to gold and therefore shows different sticking dynamics dependent on the protein's contact area with the surface. The correlation of high amplitudes with long FDHMs would then arise due to increased overlap between protein sections with high polarizability (i.e. Hem's iron-complex) and the NR's near field. 2) Hemoglobin has different affinities to different types of gold crystal facets. In this case the correlation of higher maximum amplitudes with longer FDHMs would imply that facets which offer higher affinities coincide with zones of higher near-field intensity. Which hypothesis is right or whether there is a process we have not covered will require further study beyond the scope of this work. 
We want to point out that GOx and Hem measurements shown here (Fig.2, 3 and 4) were performed with different detectors, i.e., the amplitude values given are not directly comparable. To obtain a direct comparison of perturbation amplitudes we determine the relative intensity changes $\Delta I_{rel}=\left\vert\frac{\Delta I_{max}}{\overline{I_{det}}}\right\vert$ caused by GOx and Hem from measurements performed with the same DC-coupled detector. We find average values of  $\Delta \overline{I_{rel}}=(3.4 \pm 1.5)\%$ for GOx and $(1.55 \pm 0.45)\%$ for Hem. The ratio $\Delta \overline{I_{rel}}(GOx)/\Delta \overline{I_{rel}}(Hem)= 2.2\pm1$ matches the corresponding molecular weight ratio of $2.5$ within the errors.
We would also like to put our time resolution into context: None of the events discussed above - even the longest (FDHM$\approx 100$\,\textmu s), would be resolved with other state-of-the-art nanoplasmonic assays \cite{Ament2012, Zijlstra2012, Dantham2013,Rosman2013, Baaske2014, Beuwer2015, Kim2016, Baaske2016,Wulf2016, Kim2017, Zhang2020, Siva2021}, which typically exhibit time resolutions in the order of milliseconds, i.e., see the whole extent of the traces displayed in Fig.\,2a(top) and 2b(top) as either a single or a set of few points. We think being able to resolve and analyse such short unspecific interactions opens up a whole new pathway for fast molecular fingerprinting. We envision that previously not resolvable differences in the interaction dynamics between specific proteins sub-domains and small weakly bonding receptor molecules on the NR's surface may be used as means to distinguish between proteins of similar mass. In the simplest case the average number of formed bonds determines the dwell time. This would further allow to gain structural information on analytes, especially unknown ones therefore mitigating the need for target-specific receptors. In addition physical analyte properties such as Stokes radii and molecular weight may be determined \textit{via} $\tau_{rise}$, $\tau_{fall}$ and relative amplitudes following NR calibration with a suited standard.
In conclusion, we have demonstrated the transient detection of single proteins with masses as low as $64$\,kDa traversing the subattoliter volumes spanned by plasmonic near fields during times as short as $100$\,ns and with a signal-to-noise ratio exceeding $5$.  We found good agreement between relative signal amplitudes and molecular weights. We have used our unprecedented temporal resolution to observe protein motion alongside the dynamics of unspecific protein surface interactions. We found distinctive differences in sticking behaviors of Glucose Oxidase and Hemoglobin and found initial evidence that suggests that our method may resolve rotational diffusion. We think this result offers but a glimpse of the additional information that may be gained on physical and biochemical processes on the timescales now made accessible by fast optoplasmonic detection.

\bibliography{references}

\section{Acknowledgements}
This work was supported by the Netherlands Organisation for Scientific Research (NWO) and has received funding from the European Union’s Horizon 2020 research and innovation programme under the Marie Skłodowska-Curie Grant Agreement no. 792595 (MDB).

\section{Author Contributions:}
MDB and MO conceived the idea. MDB planned the experiments. MDB and NA built the optical setup. MDB performed experiments and data analysis. DP obtained and analysed SEM micrographs. MDB and MO wrote the manuscript. All authors commented on the manuscript. 
\section{Data availability}
The data that support the findings of this study are available from the corresponding author upon reasonable request.

\section{Methods}
\subsection{Optical Setup(s):}
Here we list the components used in our measurements.
\begin{itemize}
    \setlength\itemsep{0em}
    \item[] Objective: Olympus UPLFLN100XOP
    \item[] Tube lens: Olympus Super Wide Tube Lens Unit
    \item[] Lasers: Toptica DL pro 785nm \& Cobolt Samba 532nm 
    \item[] APD: Thorlabs APD430A/M (Thorlabs) (DC-Coupled)
    \item[] APD: Helix-902-200 (Excelitas Technologies) (AC-Coupled)
    \item[] 10:90 Beamsplitter BSN11 (Thorlabs)
    \item[] Glan-Thompson Polarizer GTH10M-B (Thorlabs)
    \item[] Piezo Translator P-561.3CD (Phyisk Instrumente GmbH \& Co KG) 
    \item[] White-light source: EQ-99XFC (Energetiq)
    \item[] Spectrometer: QE-65000 (Ocean Optics)
    \item[] Reference Photodiodes 1\&2:   PDA36A2 (Thorlabs)\& HCA-S-200M (Femto)
    \item[] EOM: Amplitude Modulator AM532 (Jenoptik)
    \item[] AOM: MT110-A1-IR (AA Opto-Electronic)
    \item[] Achromatic $\lambda/2$-plate: RAC 4.2.10 (B. Halle) 
    \item[] Notch Filter: ZET532NF (Chroma) 
\end{itemize}
Traces were digitized with an oscilloscope (WaveSurfer 24MXs-B, Teledyne Lecroy) and streamed to a PC.
\subsection{Slide preparation:}
CTAB-capped gold nanorods were purchased from Nanopartz. GNR stock solutions containing $10$\,mM CTAB were sonicated (20 min./ Branson 2510) and then deposited onto glass slides (Borosilicate glass diameter $25$\,mm thickness No.1, VWR ), which were previously sonicated in ethanol and Milli-Q ($30$\,min. each/ Branson 2510), via spin-coating (Specialty coating Systems Spin Coater 6700). The CTAB-layer was consequently removed via UV-cleaning ($10-60$\,min., Ossila) and the slide was rinsed with Milli-Q water.
\subsection{Protein measurements:}
Solutions containing various concentrations of Hem and GOx were prepared  freshly before the start of each measurement. For GOx we used aqueous solutions (Milli-Q water) containing $20$\,mM of Sodium Chloride and for Hem we used $1\times $PBS (Phosphate Buffered Saline) buffer solutions.  All chemicals and proteins were purchased from Sigma-Aldrich. 
\end{document}


\title{Transient Optoplasmonic Detection of Single Proteins with Sub-Microsecond Resolution\\ Supporting Information}
\author{Martin D. Baaske}
\author{N. Asgari}
\author{D. Punj}
\author{Michel Orrit}
\affiliation{Huygens-Kamerlingh Onnes Laboratory, Leiden University, Postbus 9504, 2300 RA Leiden, The Netherlands}
\date{\today}
\maketitle
\newpage

\section{S1: Autocorrelation Measurements}
Here we show intensity autocorrelations measured for Hemoglobin and Glucose Oxidase. Autocorrelations were computed from intensity traces with $10$\,ms durations and time resolution of $10$\,ns  following the procedures used in our previous work\cite{Baaske2020}. Results are averaged over $50$ consecutively recorded traces and shown in figure\, S1. For Glucose Oxidase we find strong correlations on timescales shorter than $100$\,ns, whereas no such correlations are present for Hemoglobin. Glucose Oxidase presents the shape of two cylinders co-joined orthogonal to the cylinder axis and therefore possess a higher asymmetry then the globular Hemoglobin. Therefore we think it is likely that the difference between the autocorrelation is due to rotational diffusion. This hypothesis is further supported as we find good agreement between the observed correlation times $<100$\,ns and the rotational diffusion times $\tau_{D}$ of $28$ to $129$\,ns expected for a sphere with the dimensions of Glucose oxidase, i.e. radii of $3$ to $5$\,nm. Calculations were performed using the relation $\tau_D = (6D)^{-1}$ whereas the rotational diffusion constant $D$ was determined via the Debye-Stokes-Einstein relation: $D= \frac{k_b T }{8\pi \eta r_h^3}$, where $k_B$ is Boltzmann's constant, $T$ the temperature (here $293$\,K), $\eta$ denotes the dynamic viscosity of the medium (here water) and $r_h$ is the diffusor's hydrodynamic radius.      
\begin{figure}[htbp]
\centering
\includegraphics{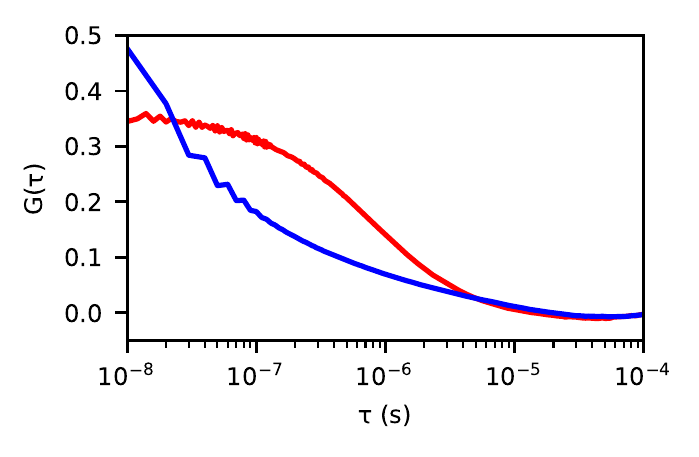}
\caption{Averaged intensity autocorrelations measured for Glucose Oxidase (blue) and Hemoglobin (red). Only the autocorrelation curve of the more asymmetric Glucose Oxidase exhibits an additional increase on timescales shorter than $100$\,ns. This increase coincides with the expected rotational diffusion times of $28$ to $129$\,ns. }
\end{figure}    
\section{S2: SEM-micrographs and sizes of gold nanorods}
Here we show the dimensions of the gold nanorod samples used in this manuscript (see Fig.\,S2). We further compare NR diameters as determined via SEM and optical means, i.e., via normalized white-light scattering spectra and polarisation angle rotation (see Fig.\,S3).  
\begin{figure}[htbp]
\centering
\includegraphics{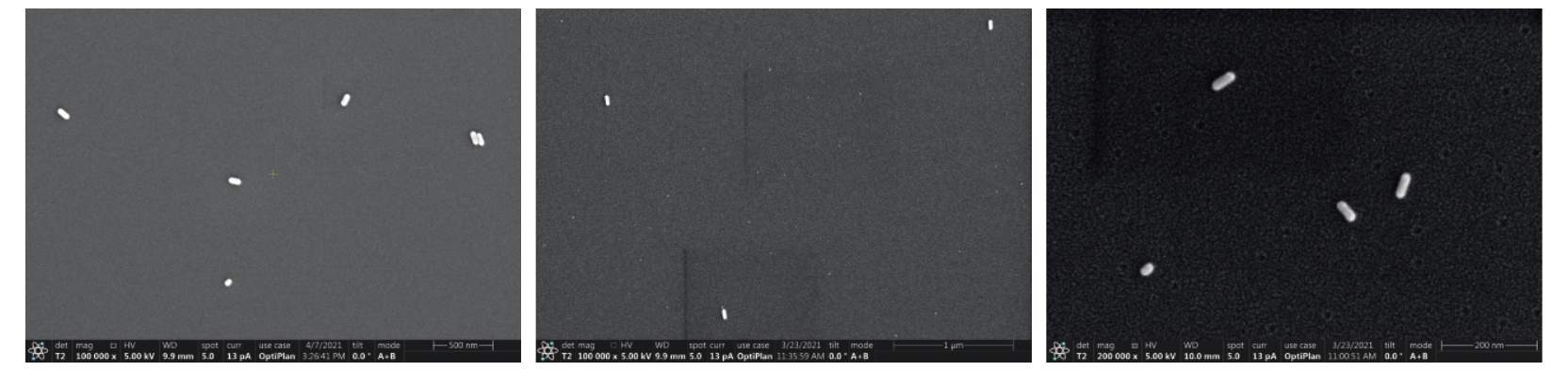}
\caption{Example SEM micrographs showing nanorod samples with average diameters of $40$, $25$, and $20$\,nm (left to right). NRs have various aspect ratios and suited candidates are selected based on white-light scattering spectra.}
\end{figure}    
\begin{figure}[hbtp]
\centering
\includegraphics{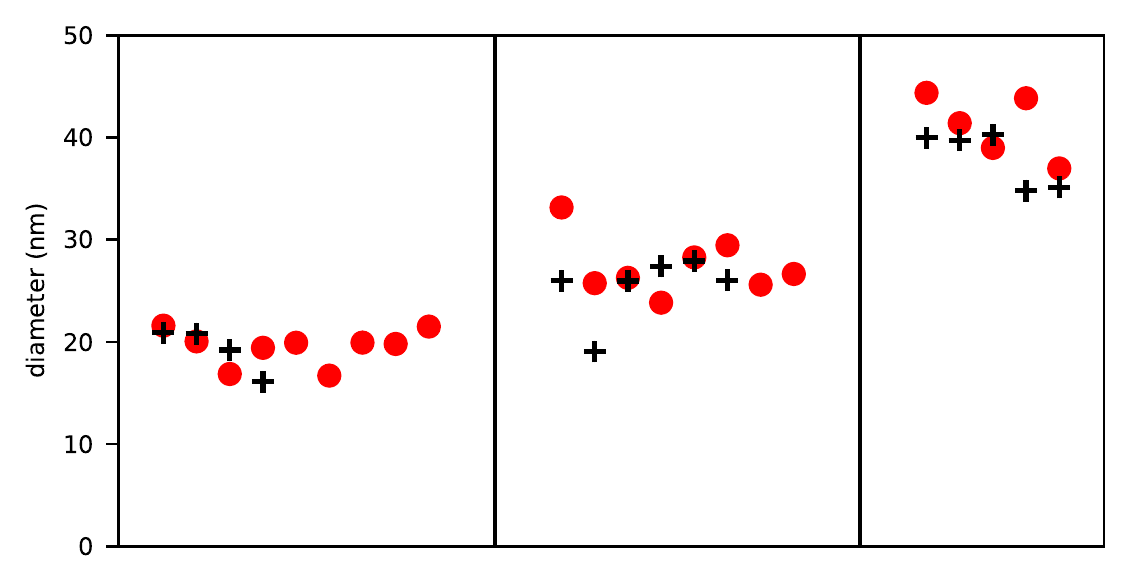}
\caption{Comparison of nanorod diameters as obtained via optical means (black, crosses) and via SEM micrographs (red, circles). Samples are not correlated, i.e., different NRs are observed via both methods. We used three different commercial NR samples, which are separated by the vertical lines. }
\end{figure}    
\bibliography{references}